\documentclass[twocolumn,preprintnumbers,amsmath,amssymb]{revtex4}

\usepackage{graphicx}
\usepackage[dvipsnames]{color}
\usepackage{bm}
\usepackage{amsmath}
\usepackage[T1]{fontenc}

\begin{document}
\def \brho{{\hbox{\boldmath $\rho$}}}
\def \beps{{\hbox{\boldmath $\epsilon$}}}
\def \bdelta{{\hbox{\boldmath $\delta$}}}

\title{Metal surface induces strong acoustic plasmons in chemically doped graphene}

\author{Vito Despoja$^{1,3}$}
\email{vito@phy.hr}
\author{Dino Novko$^{1,3}$}
\author{Ivor Lon\v cari\' c$^{2,3}$}
\author{Neven Goleni\' c$^{1}$}
\author{Vyacheslav M. Silkin$^{3,4,5}$}

\affiliation{$^1$Institute of Physics, Bijeni\v cka 46, HR-10000 Zagreb, Croatia}
\affiliation{$^2$Ru\dj er Bo\v skovi\' c Institute, Bijeni\v cka 54, HR-10000 Zagreb, Croatia}
\affiliation{$^3$Donostia International Physics Center (DIPC), P. Manuel de Lardizabal, 4, 20018 San Sebasti\'an, Spain}
\affiliation{$^4$Departamento de Fisica de Materiales and Centro Mixto CSIC-UPV/EHU, Facultad de Ciencias Qu\'{\i}micas, Universidad
del Pais Vasco UPV/EHU, Apto. 1072, 20080 San Sebasti\'an, Spain}
\affiliation{$^5$IKERBASQUE, Basque Foundation for Science, 48011 Bilbao, Spain}

\begin{abstract}
Recent theoretical considerations have demonstrated that freestanding graphene doped with alkali metals 
(AC$_x$) supports strong Dirac and weak acoustic plasmons. Here we show that when AC$_x$ is deposited on a metallic surface, the intense Coulomb 
screening completely washes out these collective modes. However, even small increase  of separation between AC$_x$ and metallic surface causes recovery of AC$_x$  plasmonic properties and 
especially the enhancement of acoustic plasmons intensities not present in the freestanding case. We further provide the physical background of these intriguing phenomena. The studied systems consist of 
lithium- and cesium-doped graphene deposited on Ir(111) surface. 
\end{abstract}

\maketitle

\section{Introduction}
Ground state crystal and electronic structure of graphene doped with alkali atoms (AC$_x$, A $=$ Li, Na, K, Cs) on different metallic surfaces, such as Ir(111), Cu(111) and Ni(111), have been recently extensively studied in several experimental and/or theoretical papers \cite{Li_Ir111_1,Li_Ir111_2,Cs_Ir111,Li_Cs_Ir111,Eu_Cs_Ir111,Li_Na_K_Cu111,Cs_Ni111}. 
The goal of these investigations was to achieve the self-standing graphene (decoupled from the surface as much as possible) with the smallest Moire corrugation.
Moreover, the mentioned experiments and further density functional theory (DFT) calculations\,\cite{dino_nl} show that graphene doped by alkali atoms possess electronic 
band structure that could potentially support very interesting plasmonic properties. 
On the one hand alkali atoms in two-dimensional (2D) superlattice metalize and form a parabolic $\sigma$ band that crosses Fermi level \cite{npj2d}. 
On the other hand, alkali atoms donate electrons to graphene $\pi$ band, lifting the Fermi level for more than 1 eV above the Dirac point \cite{dino_nl}. 
The coexistence of these partially occupied 2D bands in AC$_x$ could support at least two electronic collective modes, where one is suppose to be very strong Dirac plasmon (DP). 


The main goal of this paper is to emphasize very interesting plasmonic effects appearing in graphene doped by alkali metals. 
Recent theoretical investigations \cite{Leo1,Leo2} have already pointed out the appearance of a strong DP and a weak acoustic plasmon (AP)\,\cite{ASP1,ASP2,ASP3} in AC$_x$. 
However, here we focus on much more realistic situation where doped graphene is deposited on metallic surface \cite{NPJ,Politano,principi,echarri} and especially we aim to explore how the vicinity of metallic surface can be exploited to modify the DP and AP dispersion relations and intensities. Investigations of graphene-metal heterostructures is actually of both fundamental\,\cite{lundeberg} and practical\,\cite{iranzo,rodrigo} importance. For instance, in the field of graphene-based biosensing\,\cite{rodrigo} the graphene-metal contacts are unavoidable and thus deciphering the microscopic screening mechanisms in these structures is of great value.

In this paper we investigate the low-energy collective electronic excitations (2D plasmons) in lithium- and cesium-doped graphene (i.e., LiC$_2$ and CsC$_8$) deposited on the Ir(111) surface. 
The special attention is paid to explore how the vicinity of Ir(111) surface modifies DP and enhances AP intensities. 
We show how the self-standing doped graphene supports very strong DP and two orders of magnitude weaker AP. 
When AC$_x$ is at equilibrium separation from the Ir(111) surface, the strong metallic screening destroys the corresponding plasmonics such that the DP in LiC$_2$ becomes very weak acoustic-like branch, while the DP in CsC$_8$ almost completely disappears. 
However, small displacement from the equilibrium separation  induces the recovery of plasmon modes as well as the appearance of interesting plasmonic phenomena not present in the freestanding doped graphene. 
For instance, for displacement of  $0.6-1.2$~\AA\ very strong AP branch appears in LiC$_2$ and for displacement of $1.2-1.6$~\AA\ CsC$_8$ supports two intense AP modes. 
For larger wave vectors ($Q>0.1$\,a.u.) the intensities of these APs (laying in infra-red frequency range, $\omega>1$ eV) can be even two orders of magnitude stronger than the DPs intensities in the self-standing AC$_x$. 
These very intriguing plasmonic phenomena can be used in many plasmonics applications \cite{pl1,pl2,pl3,pl4,pl5,pl6,pl7,pl8,pl9,pl10}.

In Sec.~\ref{TotalSys} we present theoretical model for the ground state crystal and electronic structure of LiC$_2$/Ir(111) and CsC$_8$/Ir(111). 
In Sec.~\ref{DynamicalScri} we present the method for calculating the effective 2D dielectric function $\epsilon({\bf Q},\omega)$ of AC$_x$/Ir(111). 
In Sec.~\ref{Results} the results for electron-energy-loss-spectra (EELS) $-\Im[\epsilon^{-1}]$ and real part of effective 2D dielectric function $\Re[\epsilon]$ in AC$_x$/Ir(111) are presented. Finally, we provide the conclusions in Sec.~\ref{Conclusions}.

\section{Theoretical model}
\subsection{Ground state}
\label{TotalSys}
\begin{figure}[t]
\centering
\includegraphics[width=4.5cm,height=5cm]{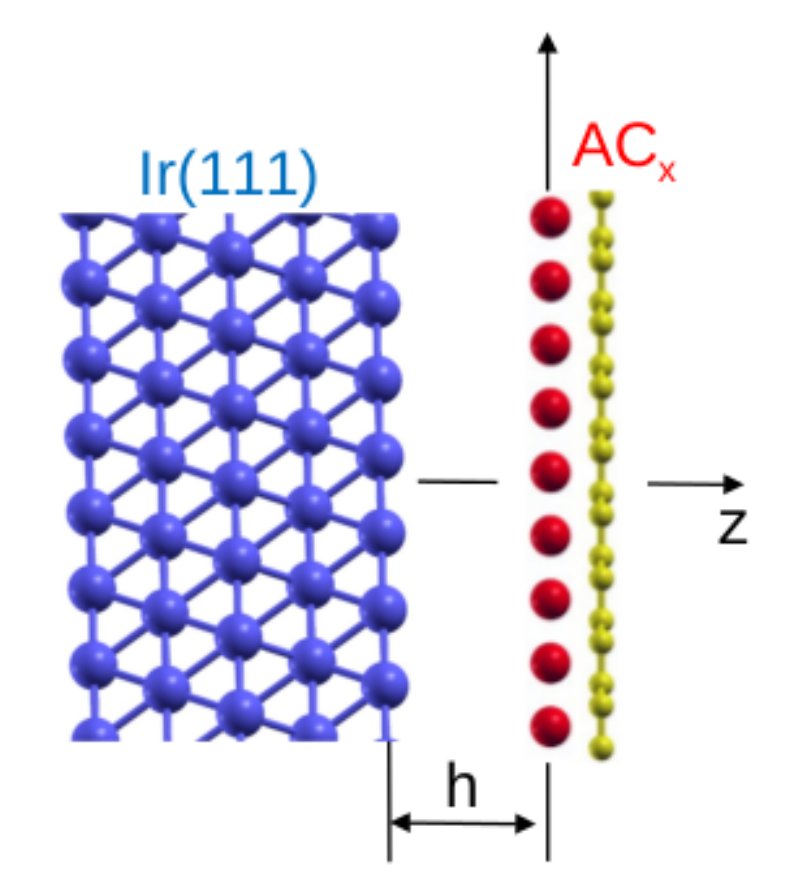}
\caption{Crystal structure of alkali atoms doped graphene on Ir(111) surface (AC$_x$/Ir(111) composite).}
\label{Fig1}
\end{figure}

The studied systems consist of graphene doped with alkali atoms and deposited on the Ir(111) surface (AC$_x$/Ir(111) composite), as shown in Fig.~\ref{Fig1}. 
The separation between alkali atom layer in AC$_x$ and topmost Ir atomic layer of Ir(111) is 
\[
h=d_{\rm Ir-A}+\Delta,
\]
where $d_{\rm Ir-A}$ represents the equilibrium separation obtained from DFT calculations and $\Delta$ represents the displacement from the equilibrium 
distance.

The crystal structure of the Ir(111) surface is modeled with 5 atomic layers. 
Graphene $2\times 2$ superlattice is matched to Ir(111)  $\sqrt{3}\times\sqrt{3}$ superlattice such that 4.8$\%$ strain is applied to Ir(111).
The unit cell in $z$ direction is set to $22$~\AA. 

Ground state electronic and structural optimization calculations were performed using the {\sc{Quantum Espresso}} (QE) package \cite{QE,QE2}. 
The core-electron interaction is approximated by the norm-conserving pseudopotentials \cite{pseudopotentials}. 
In order to capture the long range van der Waals (vdW) interaction between Ir(111) surface and doped graphene layers we use the vdW exchange correlation functional \cite{bench,irg}, in particular  vdW-DF-cx version \cite{vdW1,vdW2}. 
The ground state properties in AC$_x$/Ir(111) composites are calculated by using the $9\times9\times1$ Monkhorst-Pack K-point
mesh\,\cite{MPmesh} and the plane-wave cut-off energy is chosen to be 60\,Ry. 
The structural optimization calculations are performed until the maximum force on each atom was reduced below 0.002\,eV/\AA.
The obtained equilibrium separations between different layers in AC$_x$/Ir(111) composites are listed in Table~\ref{EqD}.
\begin{table}
\begin{tabular}{l|c|c|c|c}
\hline
\hline
\ &LiC$_2$ &CsC$_8$ & LiC$_2$/Ir(111) & CsC$_8$/Ir(111) 
\\
\hline
$d_{\rm Ir-A}$ & & &2.4&2.97
\\ 
$d_{\rm A-C}$ & 2.17 &3.0 &1.9  &3.17
\\
\hline
\hline
\end{tabular}
\caption{The equilibrium separations (in \AA) between topmost Iridium 
layer and alkali atoms layer $d_{\rm Ir-A}$ as well as between alkali atoms layer and graphene 
layer $d_{\rm A-C}$ in AC$_x$/Ir(111) composite.}
\label{EqD}
\end{table}

\subsection{Calculation of effective 2D dielectric function}
\label{DynamicalScri}
Large equilibrium distance $d_{\rm Ir-A}$ results in small electronic overlap between the AC$_x$ slab and the Ir(111) topmost layer which allows us to separate the calculation of 
the dynamically screened Coulomb interaction into two independent calculations, namely, computation of the AC$_x$ non-interacting electron response function $\chi_{\mathrm{AC}_x}^{0}$ 
and Ir(111) surface response function $D_{\rm Ir}$. Such approach considerably reduces the unit cell size and tremendously saves the computational time and 
memory requirements. This is indeed quite useful for studying the dynamical response in AC$_x$/Ir(111) composite when the AC$_x$ - Ir(111) distance is much larger than the equilibrium separation $h\gg d_{\rm Ir-A}$ or $\Delta \gg 0$. 

The ground state electronic structure of LiC$_2$ and CsC$_8$ are first calculated using the equilibrium 
positions $d_{\rm A-C}$ as in the AC$_x$/Ir(111) composite (see Table~\ref{EqD}) and using other parameters as described in Ref.\,\cite{ground}. 
The non-interacting electron response functions $\chi_{{\rm AC}_x}^{0}$ are calculated using dense K-point grids, i.e., $201\times201\times1$ and 
$101\times101\times1$ K-point meshes for LiC$_2$ and CsC$_8$, respectively. The band summations in 
$\chi_{{\rm AC}_x}^{0}$  are performed over $30$ and $100$ bands, for LiC$_2$ and CsC$_8$, respectively. In both cases the damping parameter 
$\eta=20$\,meV is used. It should be noted here that for response function calculations the crystal local field effects 
are included only in the perpendicular ($z$) direction, i.e., the response functions are 
non-local only in perpendicular direction and can be Fourier transform as  $\chi(z,z')=\frac{1}{L}\sum_{G_zG_z'}e^{iG_zz-iG'_zz'}\chi_{G_zG_z'}$ where  
$G_z$ are reciprocal space vectors in the perpendicular direction. For $\chi_{{\rm AC}_x}^{0}$ calculations we use the crystal local field energy cut-off of $10$\,Ry, which corresponds to $23$ $G_z$ wave vectors. 
The dynamically screened Coulomb interaction can be calculated by solving the Dyson equation
$w=v\  +\ v\otimes\chi_{{\rm AC}_x}^{0}\otimes w$, where $v=(2\pi/Q)e^{-Q|z-z'|}$ is bare Coulomb interaction \cite{Rukelj}, and  
$\otimes\equiv\int^{\infty}_{\infty}dz$.

\begin{figure*}[!t]
\includegraphics[width=0.3\textwidth]{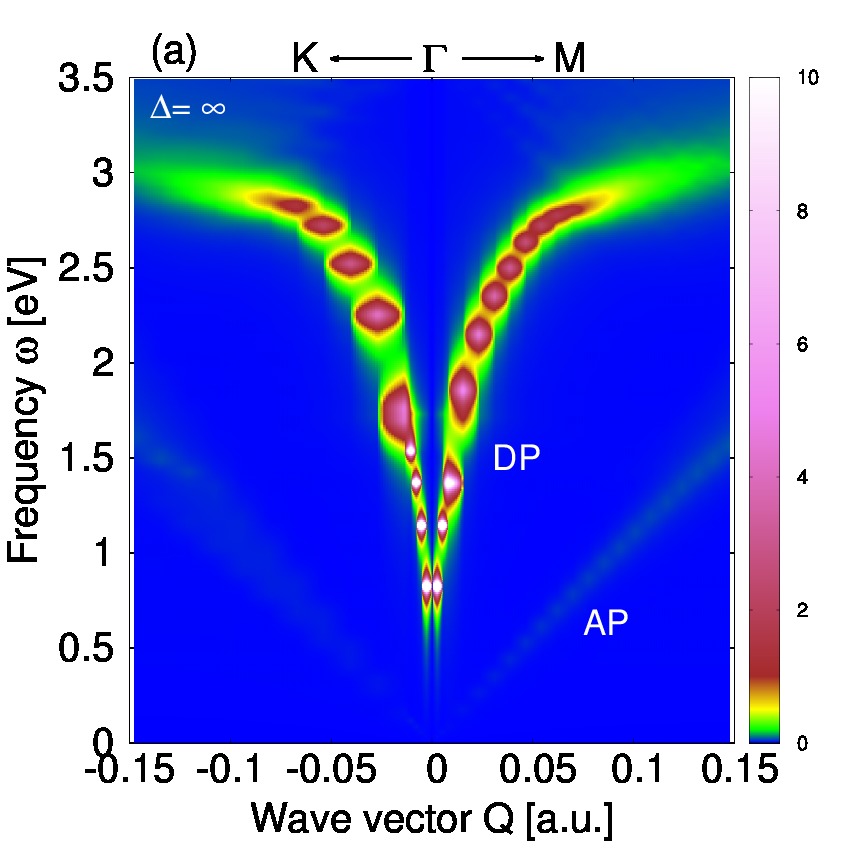}
\includegraphics[width=0.3\textwidth]{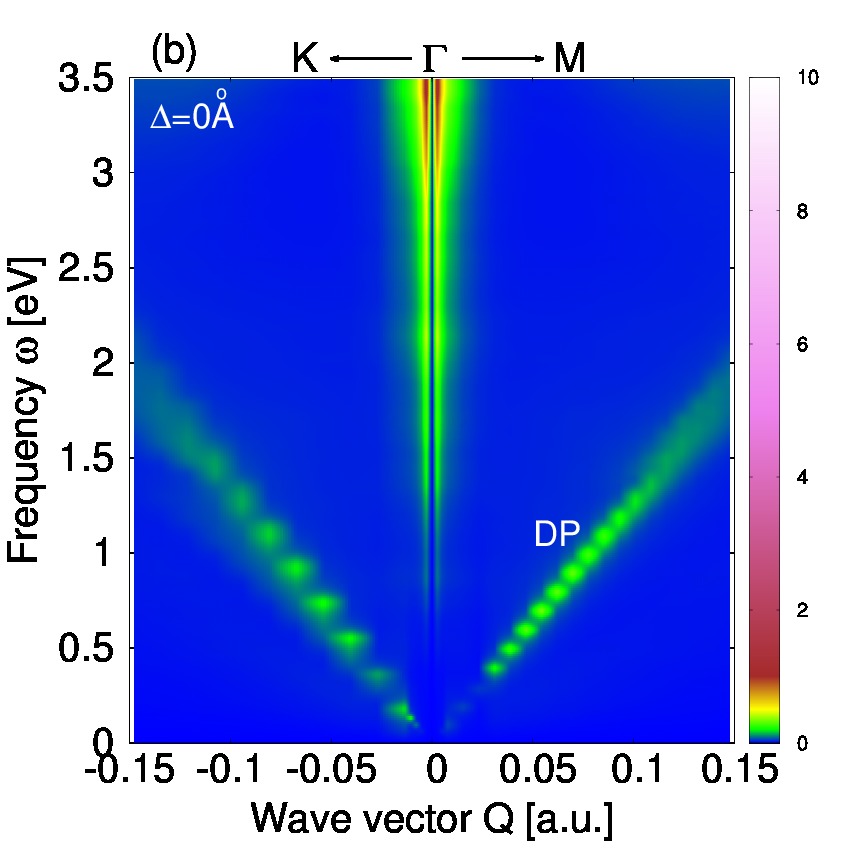}
\includegraphics[width=0.3\textwidth]{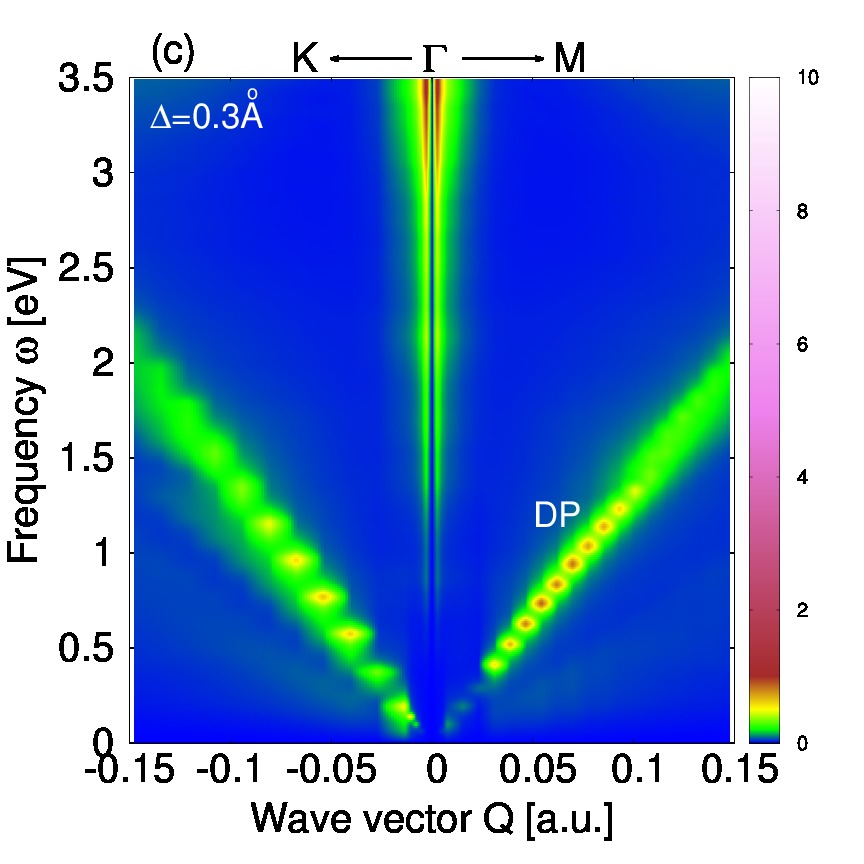}
\includegraphics[width=0.3\textwidth]{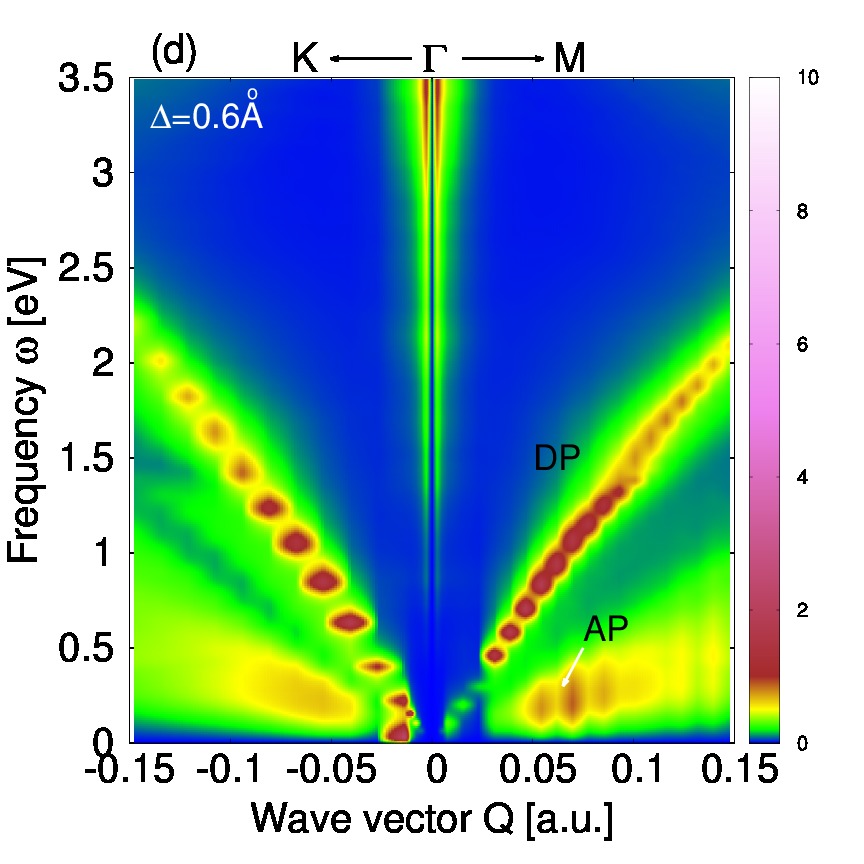}
\includegraphics[width=0.3\textwidth]{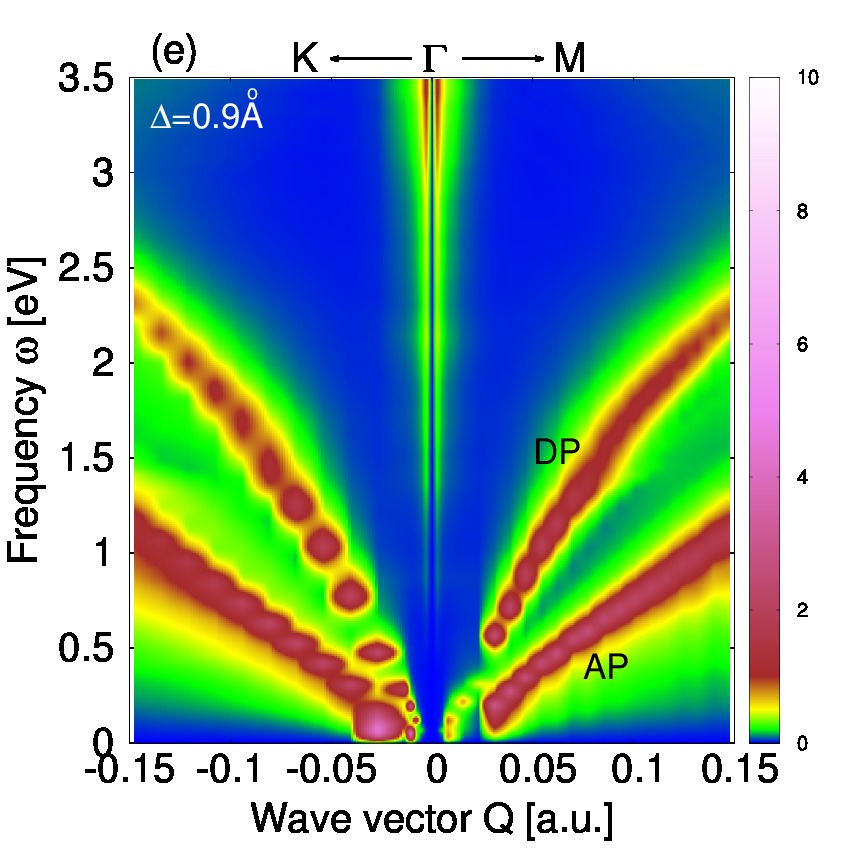}
\includegraphics[width=0.3\textwidth]{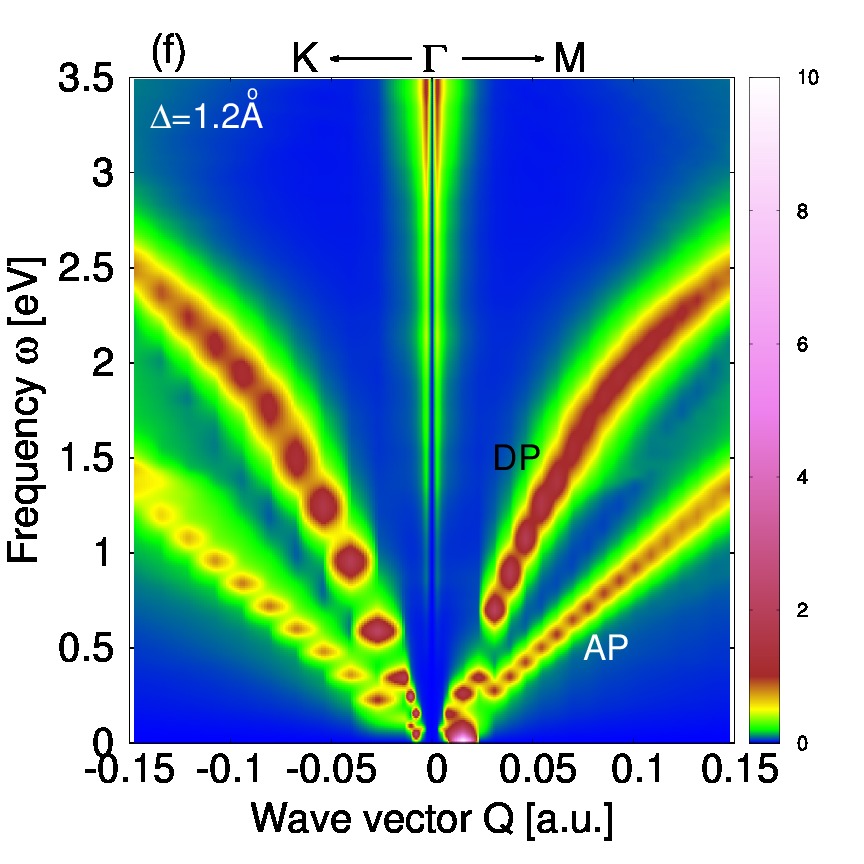}
\caption{The EELS intensity  in LiC$_2$/Ir(111) composite for  (a) $\Delta=\infty$, 
(b) $\Delta=0$, (c) $\Delta=0.3$~\AA,  (d) $\Delta=0.6$~\AA,   (e) $\Delta=0.9$~\AA,\  and (f) $\Delta=1.2$~\AA.} 
\label{Fig2}
\end{figure*}

In the vicinity  of  polarizable Ir(111) surface the Coulomb interaction (e.g., interaction between charge density fluctuations at $z,z'>-h$) is mediated by the surface screened Coulomb interaction instead of the bare interaction $v$
\[
w_{\rm Ir}\ =\ v+D_{\rm Ir}\ e^{-2Qh}e^{-Q(z+z')},
\] 
where $D_{\rm Ir}=\left\langle e^{Qz_1}\right|\chi_{\rm Ir}
({\bf Q},\omega,z_1,z_2)\left|e^{Qz_2}\right\rangle$ is the Ir(111) surface response function. The Ir(111) response 
function $\chi_{\rm Ir}$ can be obtained by solving Dyson equation $\chi_{\rm Ir}=\chi_{\rm Ir}^0+\chi_{\rm Ir}^0\otimes v\otimes\chi_{\rm Ir}$, 
where $\chi_{\rm Ir}^0$ represents the Ir(111)  non-interacting electrons response function. 
The ground state electronic structure of Ir(111)\cite{ground} surface is calculated using the $1\times 1$ unit cell. The response function $\chi_{\rm Ir}^0$ is calculated using $101\times101\times1$ K-point mesh and the 
band summations is performed over $150$ bands.
The damping parameter $\eta=30$\,meV is used. The crystal local field energy cut of $10$ Ry is used, which equal to the $37$ $G_z$ wave vectors.

After the AC$_x$ is deposited on the polarizable Ir(111) 
surface the bare Coulomb interaction $v$ has to be replaced by the surface screened Coulomb  
interaction ($v\ \ \rightarrow\ \ w_{\rm Ir}$) and the dynamically screened Coulomb interaction of the entire AC$_x$/Ir(111) composite is 
calculated by solving the ``screened'' Dyson equation:  
\begin{equation} 
w=w_{\rm Ir}\  +\  w_{\rm Ir}\otimes\chi_{{\rm AC}_x}^{0}\otimes w.
\label{screen_Dyson}
\end{equation}
Finally, the effective 2D dielectric function can be defined as  
\[
\epsilon^{-1}({\bf Q},\omega)=w({\bf Q},\omega,z=0,z'=0)/v_Q,  
\]
where $v_Q=2\pi/Q$. The EELS is then calculated  
as $S({\bf Q},\omega)=-(1/\pi)\Im[\epsilon^{-1}({\bf Q},\omega)]$.

\begin{figure*}[!t]
\includegraphics[width=0.3\textwidth]{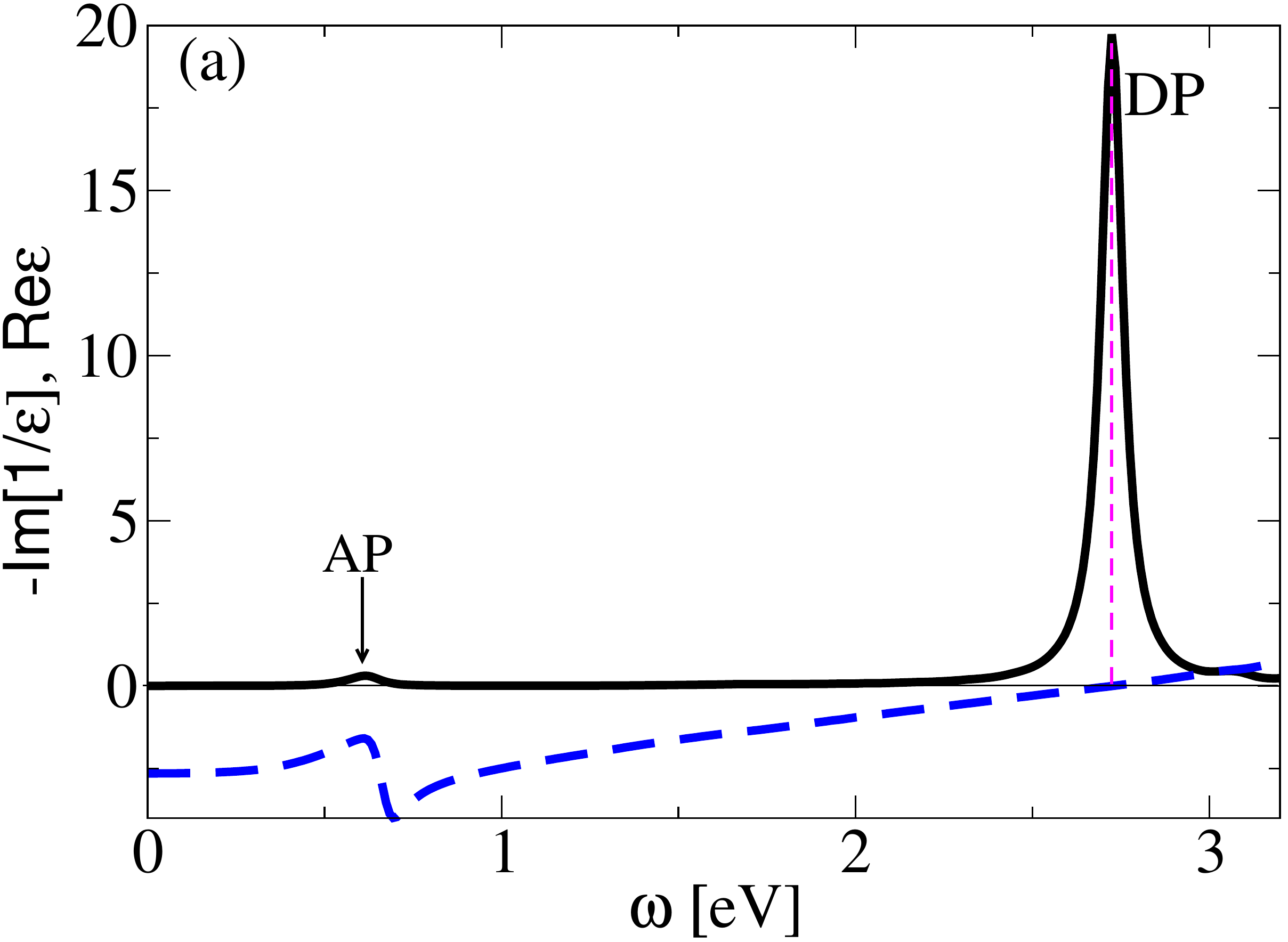}
\includegraphics[width=0.3\textwidth]{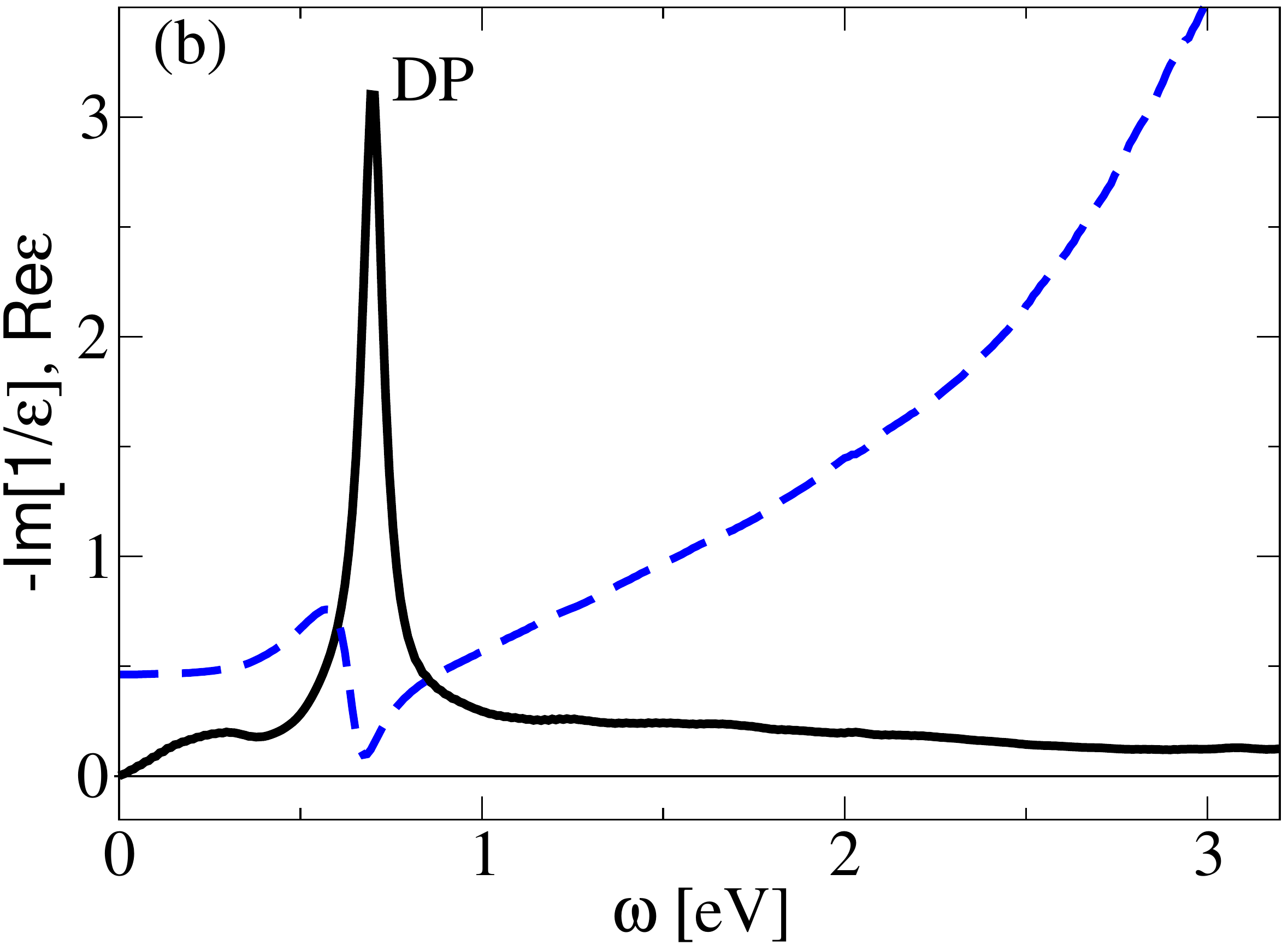}
\includegraphics[width=0.3\textwidth]{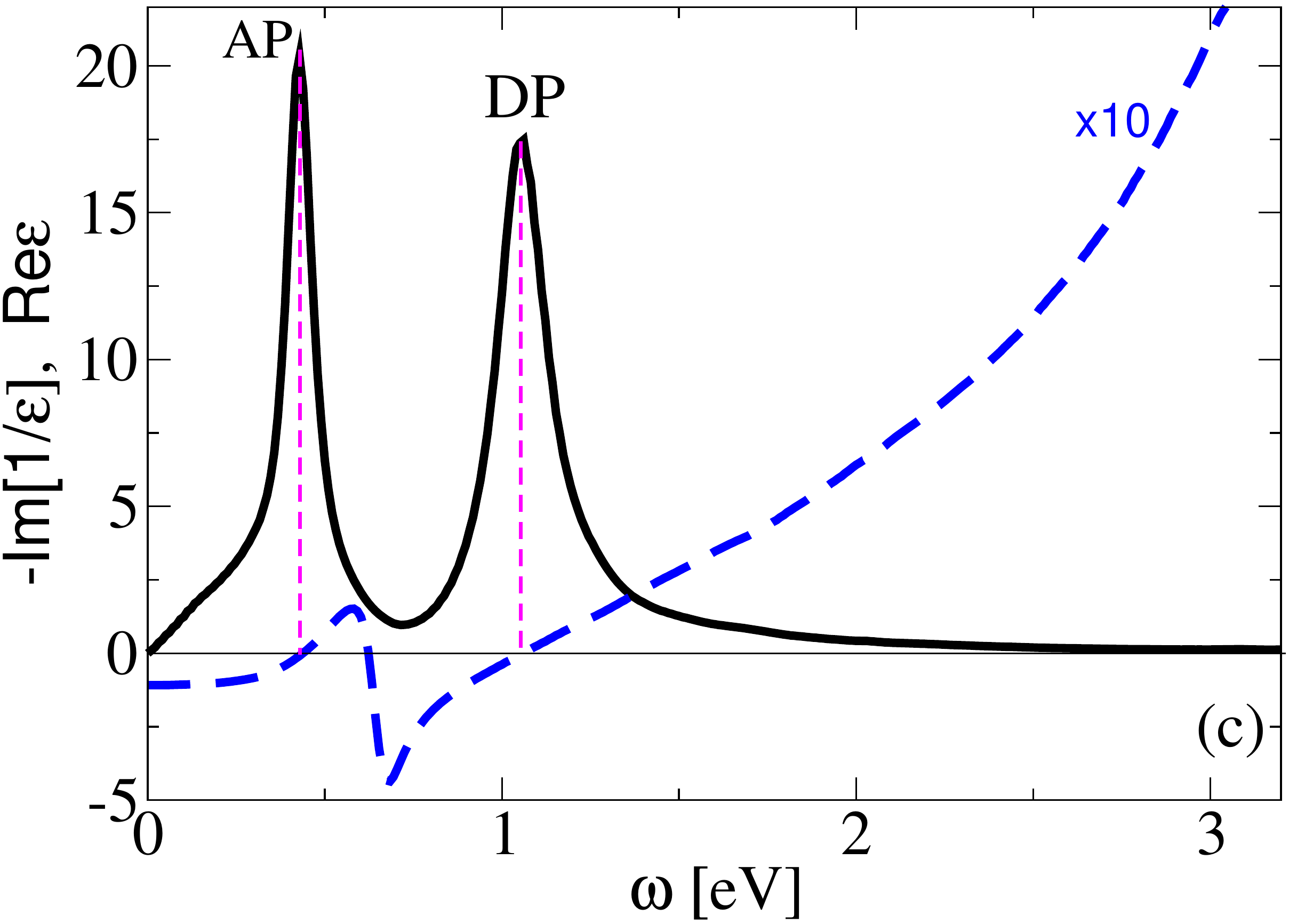}
\caption{The EELS (black solid) and $\Re[\epsilon]$ (blue dashed) in LiC$_2$/Ir(111) composite for (a) $\Delta=\infty$, 
(b) $\Delta=0$, and (c) $\Delta=0.9$~\AA. The transfer wave vector of magnitude $Q=0.054$\,a.u. is chosen to be in  $\Gamma$-M direction.} 
\label{Fig3}
\end{figure*}

\section{Results and discussion}
\label{Results}
We shall first describe the modifications obtained in the electronic excitation spectra of AC$_x$ when it is brought from  self-standing ($\Delta\rightarrow\infty$) to equilibrium ($\Delta=0$) distance. In addition, we shall gradually increase the separation $\Delta>0$, present the corresponding spectra and explain the physical background of the spectral alterations.

Figure \ref{Fig2}(a) shows the EELS intensity in self-standing ($\Delta=\infty$) LiC$_2$. As already reported in Ref.\,\cite{Leo1} in LiC$_2$ Li atoms donate electrons to graphene $\pi^*$ band and at the same time they metalize and form parabolic $\sigma$ band which remains partially filled. 
Therefore, there are two bands crossing Fermi level which result in two plasmons in LiC$_2$, i.e., strong DP and weak AP [see Fig.\,\ref{Fig2}(a)]. 
Figure \,\ref{Fig2}(b) shows EELS intensities in LiC$_2$ when it is at the equilibrium distance ($\Delta=0$) from the Ir(111) surface. 
It turns out that metallic surface radically modifies the intensities of the electronic modes in LiC$_2$. 
Namely, the DP becomes very weak and changes to a linearly dispersive mode, while the AP disappears. 
Strong metallic screening obviously significantly reduces the intensity of electronic modes and pushes them toward lower energies. 
In order to verify this claim in Figs.\,\ref{Fig2}(c)-(e) we show the EELS spectra of LiC$_2$, where the separation between LiC$_2$ and Ir(111) is gradually increased, from $\Delta=0.3$~\AA\  to $\Delta=1.2$~\AA.
Figure \ref{Fig2}(c) shows that even small displacement of $\Delta=0.3$~\AA\ from equilibrium causes substantial increase of DP intensity, while the dispersion remains linear. 
For $\Delta=0.6$~\AA\  shown in Fig.\,\ref{Fig2}(d) the broad AP appears in the lower energy region and the DP becomes stronger. 
Already for this separation the intensity of AP is an order of magnitude stronger than the intensity of the AP in  the self-standing LiC$_2$ [see Fig\,\ref{Fig2}(a)]. 
For $\Delta=0.9$~\AA\ shown in Fig.\,\ref{Fig2}(e) the dispersion relation of the DP bends (i.e., it is no longer linear) and its energy and intensity increase.   
The AP becomes strong, sharp and well defined plasmon mode with intensity two orders of magnitude larger than in self-standing LiC$_2$ and even stronger than the corresponding DP. 
For $\Delta=1.2$~\AA, shown in Fig. \ref{Fig2}(f), the AP intensity reduces for an order of magnitude (in comparison with the $\Delta=0.9$~\AA\ case) and the intensity of DP continues to increase toward its self-standing ($\Delta=\infty$) value.
It is important to note that since the metallic screening reduces the plasmon frequency, it pushes the DP out of the interband $\pi\rightarrow\pi^*$ continuum \cite{gr2013}, which in turn reduces the Landau damping and increases the intensity of the plasmon compared to the self-standing case (e.g., compare the intensities in Figs.\,\ref{Fig2}(e) and \ref{Fig2}(f) with the intensities in Fig.\,\ref{Fig2}(a) when $Q>0.1$\,a.u.).
Also, we note that for the shown wave-vector interval the EELS intensities are almost isotropic, i.e. $S({\bf Q}_{\rm \Gamma M},\omega)\approx S({\bf Q}_{\rm \Gamma K},\omega)$. 

The panels in Fig.\,\ref{Fig2} clearly show that when LiC$_2$ - Ir(111) distance is out of equilibrium $\Delta>0$, the Ir(111) surface induces the strong AP but only for small interval of separations $0.5<\Delta<1.0$~\AA. 
In order to clarify this phenomenon we show in Fig.\,\ref{Fig3} the real part of the effective 2D dielectric function (blue dashed lines) for three characteristic separations (a) $\Delta=\infty$, (b) $\Delta=0$, and (c) $\Delta=0.9$~\AA~and
for $Q=0.054$\,a.u. in the $\Gamma$-M direction.
The corresponding EELS are shown by black solid lines. 
It can be seen that in all of the cases the $\Re[\epsilon]$ contains ``kink'' structures which, depending on separation from the Ir(111) surface, shifts up and down as well as crosses zero at different frequencies.
For $\Delta=\infty$ the ``kink'', where the weak AP appears, is entirely below 
zero. 
At the higher frequencies, i.e., after the ``kink'', the $\Re[\epsilon]$ increases and crosses the zero where strong DP appears, classifying the DP as a well-defined collective electronic mode. 
On the other hand, when $\Delta=0$ the ``kink'' is entirely above the zero and $\Re[\epsilon]$ does not cross the zero at all. 
This results in the appearance of the weak DP around the dip of $\Re[\epsilon]$. 
Considering that when $\Delta=\infty$ changes to $\Delta=0$ the ``kink'' in $\Re[\epsilon]$  transforms from entirely below to entirely above the zero, there should be some $\Delta$ interval when the ``kink'' crosses zero.    
As can been seen from Fig.\,\ref{Fig3}(c) this situation occurs, e.g., when $\Delta=0.9$~\AA~resulting in the appearance of AP and DP. 
In that case both plasmons can be classified as strong 
well-defined collective modes, as is obvious from Fig.\,\ref{Fig2}(e). For slightly larger separation, i.e., $\Delta>1.0$, the ``kink'' falls entirely below the zero and the AP quickly weakens. 
However, $\Re[\epsilon]$ continues to cross zero at larger frequencies where strong DP appears, as is also obvious from Fig.\,\ref{Fig2}(f).

\begin{figure*}[!ht]
\includegraphics[width=0.3\textwidth]{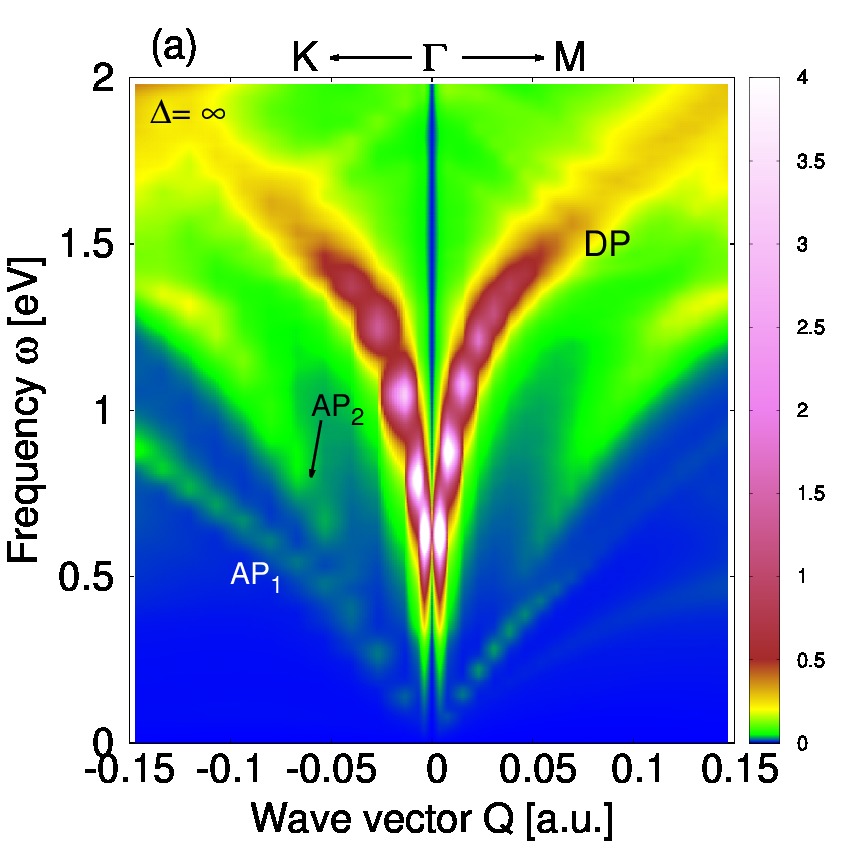}
\includegraphics[width=0.3\textwidth]{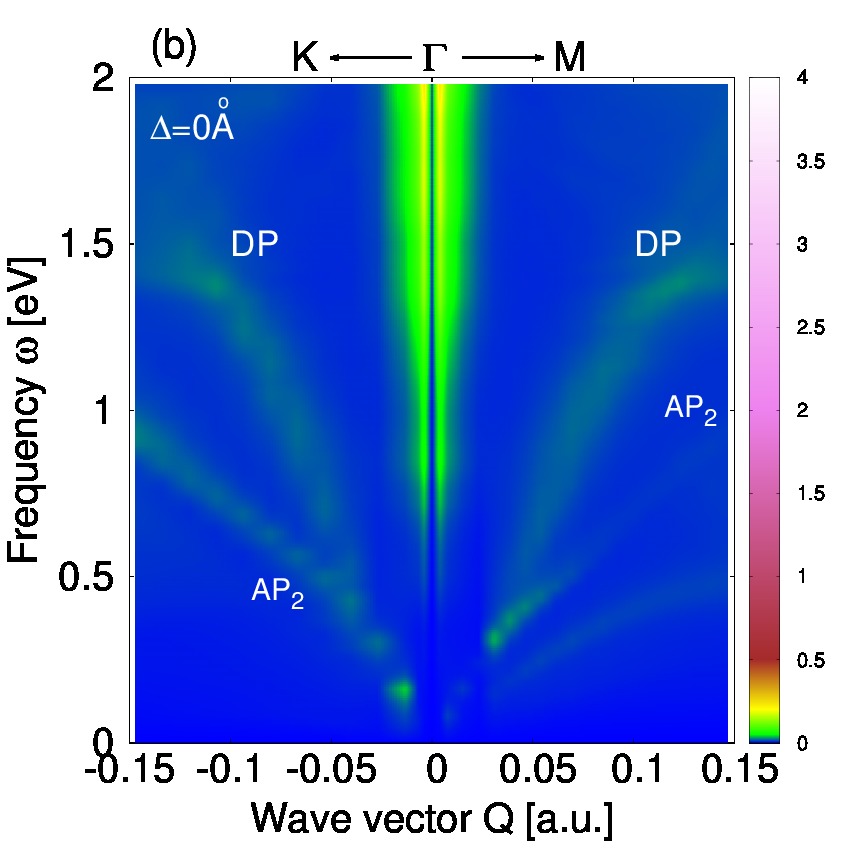}
\includegraphics[width=0.3\textwidth]{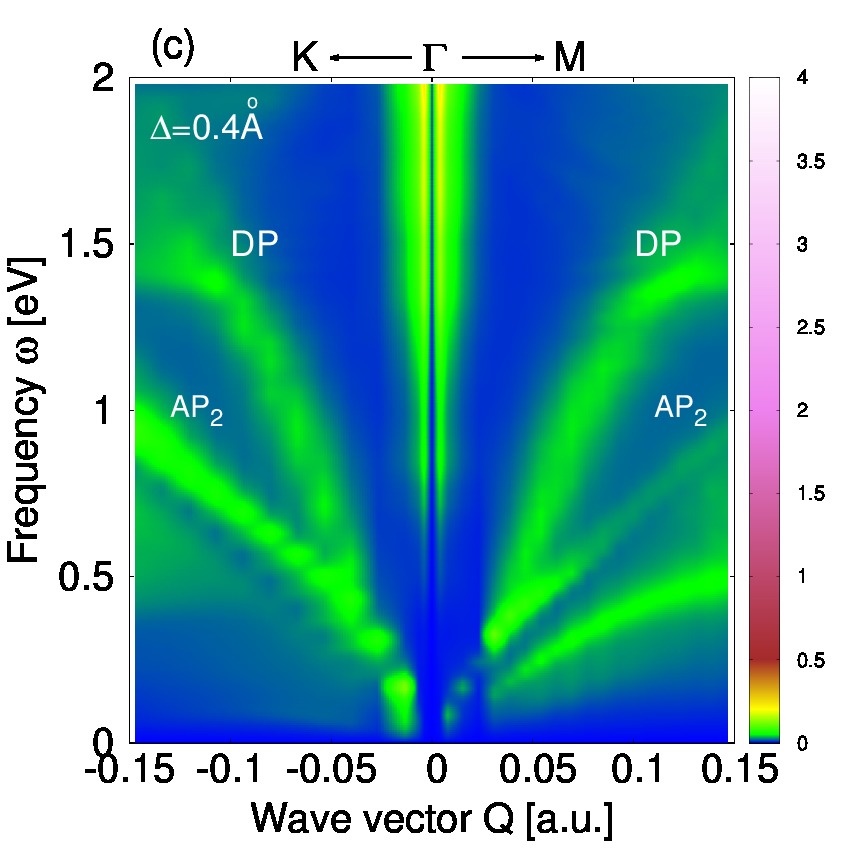}
\includegraphics[width=0.3\textwidth]{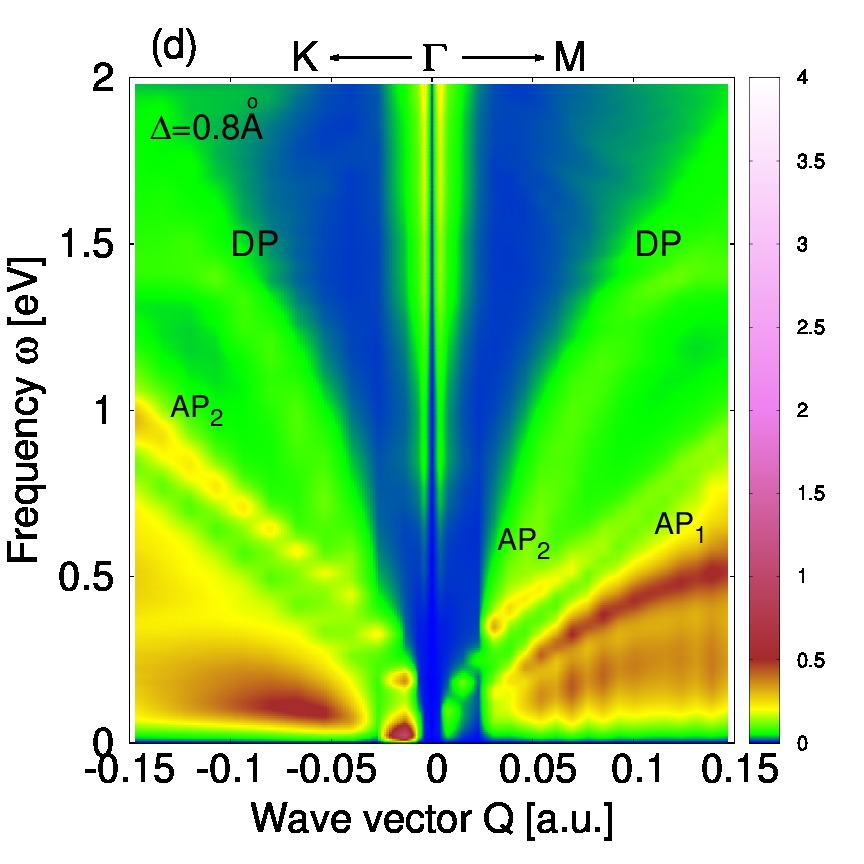}
\includegraphics[width=0.3\textwidth]{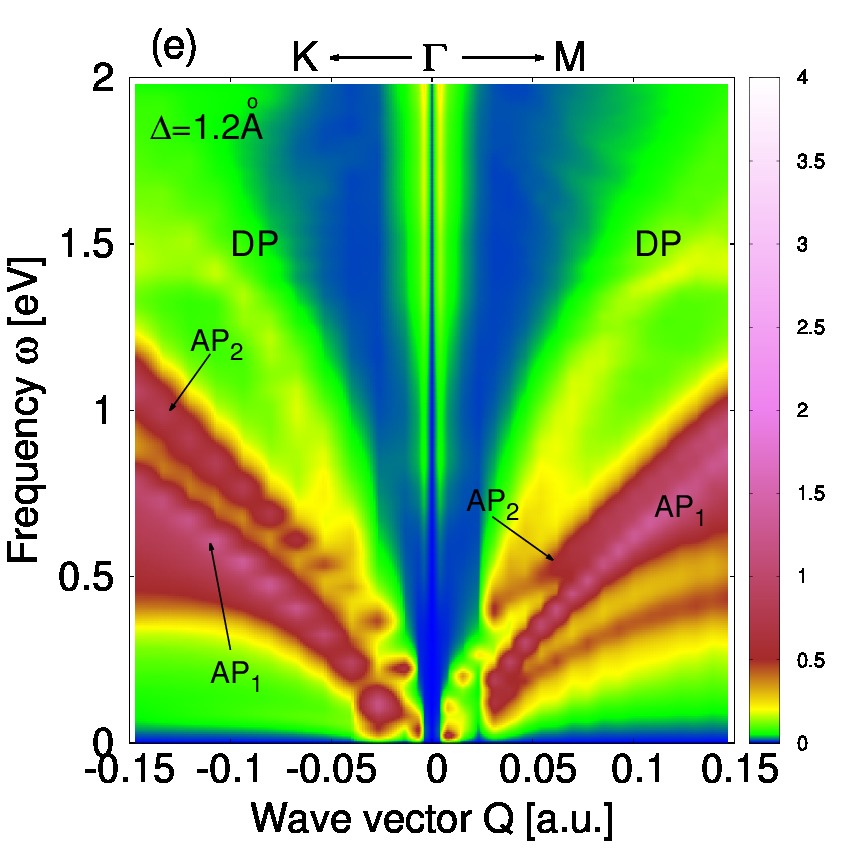}
\includegraphics[width=0.3\textwidth]{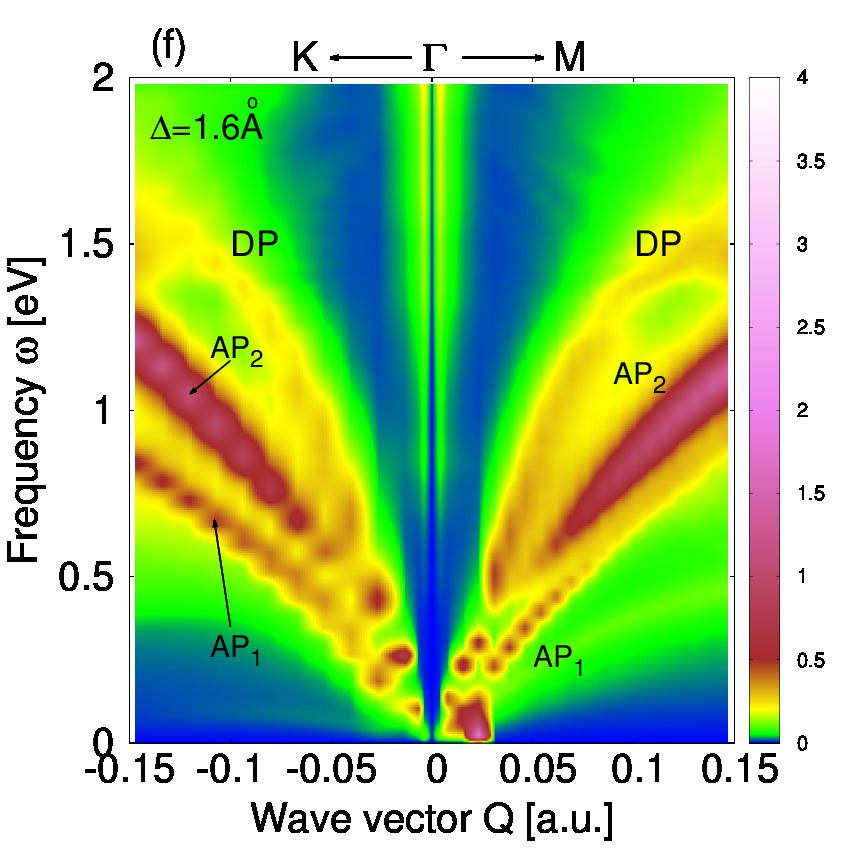}
\caption{The EELS intensity  in CsC$_8$/Ir(111) composite  for  (a) $\Delta=\infty$, 
(b) $\Delta=0$, (c) $\Delta=0.4$~\AA,\  (d) $\Delta=0.8$~\AA,\   (e) $\Delta=1.2$~\AA,\ and 
(f) $\Delta=1.6$~\AA.} 
\label{Fig4}
\end{figure*}
\begin{figure*}[!ht]
\includegraphics[width=0.3\textwidth]{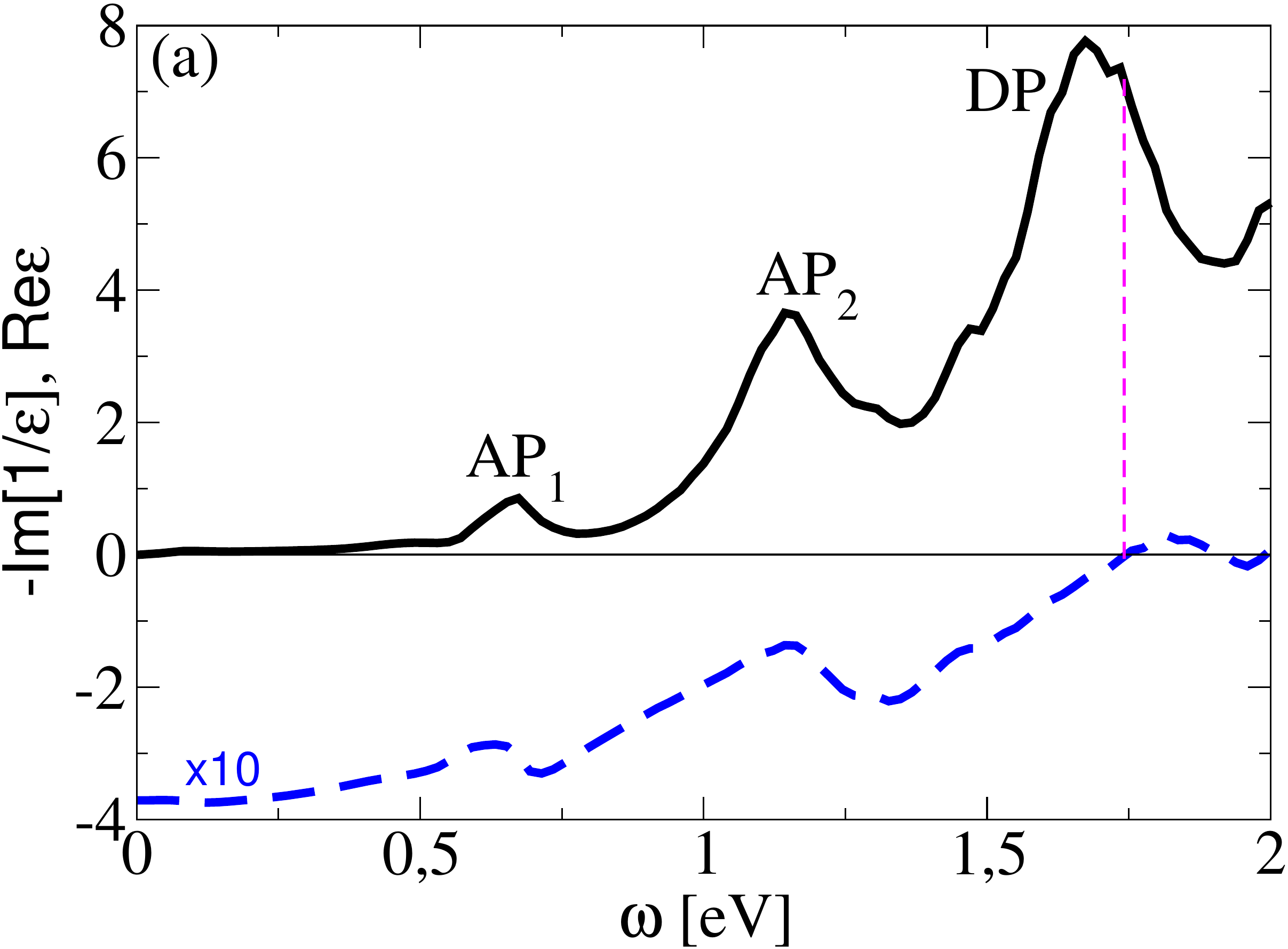}
\includegraphics[width=0.3\textwidth]{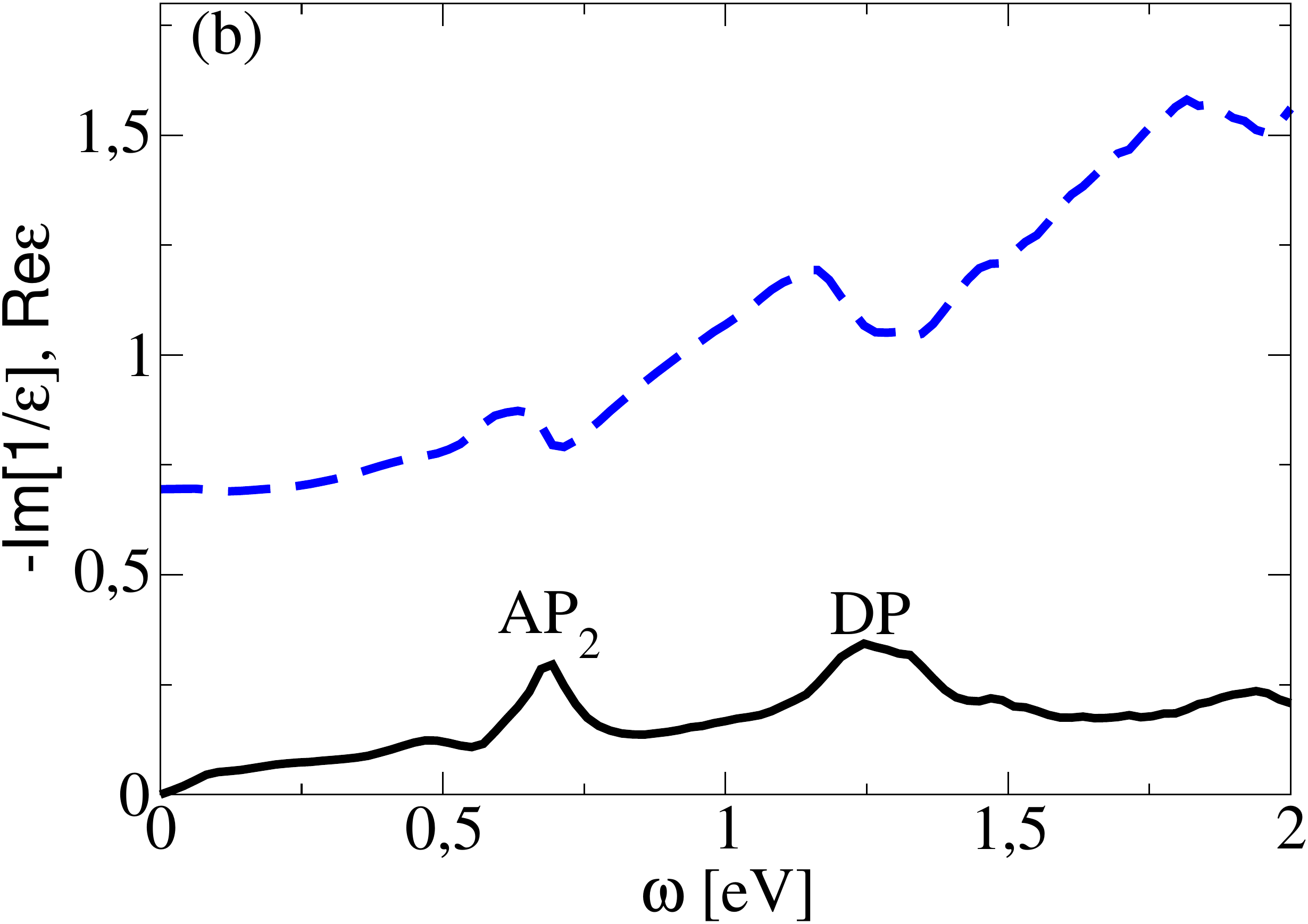}
\includegraphics[width=0.3\textwidth]{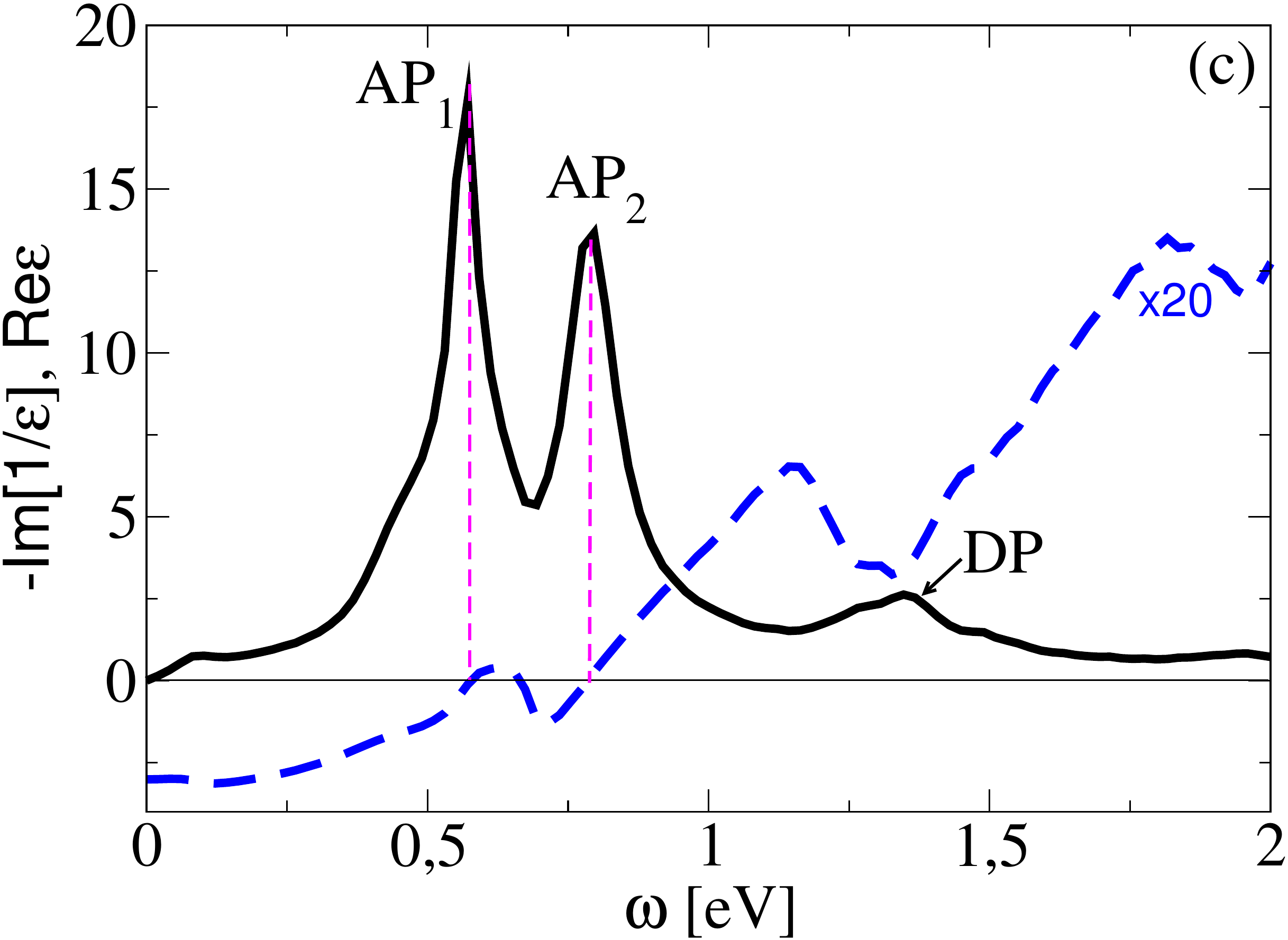}
\caption{The EELS (black solid) and $\Re[\epsilon]$ (blue dashed) in CsC$_8$/Ir(111) composite for (a) $\Delta=\infty$, (b) $\Delta=0$,  and (c) $\Delta=1.3$~\AA. The transfer wave vector of 
magnitude  $Q=0.094$\,a.u. is chosen to be in  $\Gamma$-K direction.} 
\label{Fig5}
\end{figure*}

Further, we show the EELS intensity of the self-standing ($\Delta=\infty$) CsC$_8$ in Fig.\,\ref{Fig4}(a). 
In CsC$_8$ the graphene $\pi^*$ band crosses Fermi level
twice which together with partially filled parabolic Cs($\sigma$) 
band \cite{Leo2} is sufficient for the appearance of the three plasmons, i.e., the strong DP and the two weak acoustic plasmons, which we dub AP$_1$ and AP$_2$. This nomenclature for acoustic plasmons is just formal since their dispersion will not be necessary linear.
Figure \ref{Fig4}(b) shows the EELS intensity in CsC$_8$ when it is at 
equilibrium distance ($\Delta=0$) from the Ir(111) surface. The strong metallic screening extremely weakens the oscillatory strengths of the electronic modes such 
that the DP and AP$_2$ become weak and barely visible modes, while the AP$_1$ disappears. Figures \ref{Fig4}(c)-(e) show the EELS intensities in CsC$_8$ when the separation between CsC$_8$ and Ir(111) surface gradually increases, from $\Delta=0.4$~\AA~to $\Delta=1.6$~\AA. For $\Delta=0.4$~\AA\ shown in Fig.\,\ref{Fig4}(c) the EELS intensity slightly increases, however, 
for $\Delta=0.8$~\AA\ shown in Fig.~\ref{Fig4}(d) the enhancement of the EELS intensity is already significant and new AP$_1$ plasmon appears. 
It can be noticed that AP$_1$ and AP$_2$ have square-root dispersions, which is especially visible in $\Gamma$-M direction. In $\Gamma$-K direction, the AP$_1$ is still very broad emerging mode.
For $\Delta=1.2$~\AA~shown in Fig.~\ref{Fig4}(e) the AP$_1$ and AP$_2$ turn out to be strong well-defined plasmon modes. In $\Gamma$-K direction the AP$_1$ 
and AP$_2$ have well separated dispersion relations, while in $\Gamma$-M direction and for larger wave vectors $Q>0.1$\,a.u. these modes are degenerate. 
Interesting feature occurs in long-wavelength limit $Q\approx 0$, especially in $\Gamma$-M direction, where AP$_2$ and DP start as a one plasmon branch and then at $Q\approx 0.025$a.u. they bifurcate, i.e., the DP continues as square-root and AP$_2$ as linear branches. For $\Delta=1.6$~\AA~shown in Fig.\,\ref{Fig4}(f) the AP$_1$ is already notably weakened, the AP$_2$ is still very strong mode, especially for larger wave 
vectors $Q>0.1$\,a.u. in both ($\Gamma$-M and $\Gamma$-K) directions, while the DP continues to strengthen. The observable bifurcation into AP$_2$ and DP branches is still present, in both directions for $Q\approx 0.05$\,a.u. As already discussed, in the LiC$_2$ case, for 
 $\Delta>2$~\AA~the AP$_1$ and AP$_2$ suddenly weaken and DP becomes gradually stronger until its intensity reaches the freestanding ($\Delta=\infty$) value.

In order to understand the appearance of the strong acoustic plasmons AP$_1$ and AP$_2$  in 
CsC$_8$ for certain separations $\Delta$, in the panels of Fig.\,\ref{Fig5} we show the 
real parts (blue dashed lines) of the effective 2D dielectric function ($\Re[\epsilon]$) for three characteristic 
separations (a) $\Delta=\infty$, (b) $\Delta=0$ and (c) $\Delta=1.3$~\AA, and for $Q=0.094$\,a.u in the $\Gamma$-K direction. The EELS are shown by the black solid lines. 
Contrary to the LiC$_2$ case, it can be seen that the $\Re[\epsilon]$ of CsC$_8$ has two ``kinks'' which move up and down 
depending on the separation $\Delta$. This means that for certain separations $\Delta$ the $\Re[\epsilon]$ can cross zero three times, and, therefore, three well-defined plasmons appear. Figure \ref{Fig5}(a) shows that both ``kinks'' are entirely below zero, which gives only weak plasmons. 
For larger frequencies, after the ``kink'' structures, the $\Re[\epsilon]$ crosses zero providing relatively strong, but quite broad DP. This is reasonable considering that for this wave 
vector the DP already enters $\pi\rightarrow\pi^*$ interband continuum and is thus considerably Landau damped\cite{gr2013}. This effect can also be seen as sudden decrease of DP 
intensity for $Q>0.05$\,a.u. in Fig.\,\ref{Fig4}(a). For $\Delta=0$ the ``kinks'' are entirely above the zero and $\Re[\epsilon]$ does not cross
the zero at all. This results in only two very weak plasmons AP$_2$ and DP around where the dips of the ``kink'' structures appear.
Mapping the conclusions from the LiC$_2$ case, we expect a narrow $\Delta$ interval as well for the CsC$_8$ where at least one of the ``kinks'' crosses zero.
It should also be noted that for this system there is no $\Delta$ for 
which both ``kinks'' would cross zero simultaneously. 
Figure \ref{Fig5}(c) shows that for $\Delta=1.3$~\AA~the first ``kink'' crosses zero 
two times which results in strong and well-defined plasmons AP$_1$ and AP$_2$. The second ``kink'' is still above zero resulting in a weak 
DP just below the dip in the ``kink''. This  corresponds to the situation shown in Fig.\,\ref{Fig4}(e) where the two strong plasmon, i.e., AP$_1$ and AP$_2$, and the weak DP bands are present.
It is important to note here that for $\Delta=1.3$~\AA\ [Fig.\,\ref{Fig5}(c)] the intensities of AP$_1$ and AP$_2$ are an order of magnitude larger than AP$_1$ and AP$_2$ intensities and 
more than two times larger than DP intensity in CsC$_8$ for equilibrium separation $\Delta=0$ [Fig.\ref{Fig5}(c)], at the same wave vector $Q$.  
For slightly larger $\Delta>2.0$~\AA~both 'kinks' fall below the zero and AP$_1$ and AP$_2$ quickly weaken, however the $\Re[\epsilon]$ continues to cross zero for larger frequencies where stronger DP is present (not shown).

These results generally show that for narrow interval of displacements [i.e.,  $0.5$~\AA$\leq\Delta\leq 1.0$~\AA~for LiC$_2$/Ir(111) and $1.0$~\AA$\leq\Delta\leq 1.5$~\AA~for CsC$_8$/Ir(111)] 
the  AC$_x$/Ir(111) composites support strong APs which are for small wave vectors (i.e., $Q<0.05$\,a.u.) as strong as the DP in the self-standing cases. These APs persist as strong well-defined collective 
modes in the wide wave vector interval (i.e., $0<Q<0.15$\,a.u.) such that for larger wave vectors ($Q>0.1$ a.u.) APs may be even three orders of magnitude stronger [see Figs.\,\ref{Fig2}(e)-(f) and Figs.\,\ref{Fig4}(e)-(f)] than the Landau damped DPs in self-standing systems [see Fig.\,\ref{Fig2}(a) and Fig.\,\ref{Fig4}(a)]. The metallic screening pushes the DP towards lower 
frequencies such that in the LiC$_2$/Ir(111) composite DP leaves the interband $\pi\rightarrow\pi^*$ continuum and becomes sharp long-lived mode.
Achieving these very interesting plasmonic features requires an artificial ejection of the AC$_x$/Ir(111) from the equilibrium distance $\Delta$ which might be experimentally feasible. For instance, by placing the inert single or few layers of hexagonal boron-nitride between graphene and metal surface, as it is actually the case in the contemporary graphene-based devices\,\cite{lundeberg,iranzo}. Or by replacing the alkali atoms by more larger intercalants, such as the FeCl$_3$ molecules that act as the electron acceptors lowering the Fermi energy below $-1$\,eV\,\cite{bezares}. Furthermore, the  desirable plasmonic properties could be even achieved by changing the concept of the composite system. For example, 
instead of using metallic substrate we can use semiconducting (e.g., SiO$_2$) substrates, while the plasmonic properties in AC$_x$/substrate could be manipulated by the dull metallic (e.g., aluminum) tip from the above.

\section{Conclusions}
\label{Conclusions}
We have demonstrated how the 2D collective modes in chemically doped graphene are drastically modified in the presence of the Ir(111) metallic surface. For instance, very strong Dirac plasmon present in LiC$_2$ becomes two order of magnitude weaker and linearly dispersive, while the acoustic plasmon disappears. Similar modifications were as well observed in CsC$_8$.
Further, to understand this physical phenomena we gradually increased the graphene-surface separations $\Delta$ 
and analyzed the effective 2D dielectric function $\epsilon$. 
We have found that for the equilibrium separation ($\Delta=0$) the strong metallic screening  pushes the real part of effective 2D dielectric function 
($\Re[\epsilon]$) entirely above the zero which blocks the formation of well-defined 
2D collective modes. However, for narrow interval of out-of-equilibrium separations ($\Delta>0$) the $\Re[\epsilon]$ crosses zero twice providing two strong 2D plasmons. In particular,
for $0.5$~\AA$\leq\Delta\leq 1.0$~\AA~the LiC$_2$/Ir(111) composite supports strong well-defined Dirac and acoustic plasmons, while for  $1.0$~\AA\,$\leq\Delta\leq 1.5$~\AA~the 
CsC$_8$/Ir(111) composite  contains two intense acoustic plasmons when the wave vector is smaller than $0.15$\,a.u.
All in all, we have shown that the chemically doped graphene in the presence of the metallic surface could support even superior plasmonic features compared to the freestanding case.
This manipulative plasmonic property can be used in many applications such as biosensing or in plasmon enhanced spectroscopic techniques. 


\section*{Acknowledgment}
V.D.\ acknowledges support from the QuantiXLie Center of Excellence, a project cofinanced by the Croatian Government and European Union through the European Regional Development Fund - the Competitiveness and Cohesion Operational Programme (Grant No.\ KK.01.1.1.01.0004). V.D. is also grateful to Donostia International Physics Center (DIPC) for hospitality during various stages of this work.
D.N. acknowledges financial support from the European Regional Development Fund for the ``Center of Excellence for Advanced Materials and Sensing Devices'' (Grant No. KK.01.1.1.01.0001).
I.L. acknowledges support from the European Union through the European Regional Development Fund - the Competitiveness and Cohesion Operational Programme (KK.01.1.1.06).
Computational resources were provided by the DIPC Computing Center.


\begin{thebibliography}{99}
\bibitem{ground} DFT calculations of separate layers are performed in the same way as for the complete system, except XC potential in LiC$_2$ is approximated by the Perdew-Zunger local density approximation (LDA) \cite{LDA} and in CsC$_8$ and Ir(111) surface  by Perdew-Burke-Ernzerhof generalized gradient approximations (GGA)\cite{GGA}. 
We used $12\times12\times1$, $6\times6\times1$ and $12\times12\times1$ Monkhorst-Pack K-point mesh for LiC$_2$, CsC$_8$, and Ir(111), respectively.
Supercells in $z$ direction for AC$_x$ are set to 12.3~\AA~and for Ir(111) surface to 20.9~\AA.
\end{thebibliography}
\end{document}